\documentstyle[preprint,eqsecnum,aps]{revtex}
\tightenlines

\begin{document}
\draft
\preprint{number}
\title{Exact Solutions of Five Dimensional Anisotropic Cosmologies}
\author{Paul Halpern}
\address{Department of Mathematics, Physics and Computer Science\\
University of the Sciences in Philadelphia\\
600 S. 43rd Street\\
Philadelphia, Pa. 19104}
\date{March 15, 2002}
\maketitle
\begin{abstract}
We solve the five dimensional vacuum Einstein equations for several kinds of anisotropic geometries.  We consider metrics in which the spatial slices are characterized as Bianchi types-II and V, and the scale factors are dependent both on time and a non-compact fifth coordinate.  We examine the behavior of the solutions we find, noting for which parameters they exhibit contraction over time of the fifth scale factor, leading naturally to dimensional reduction.  We explore these within the context of the induced matter model: a Kaluza-Klein approach that associates the extra geometric terms due to the fifth coordinate with contributions to the four dimensional stress-energy tensor. 
\end{abstract}

\narrowtext

\section{Introduction}
\label{sec:intro}

Recent advances in string and membrane theories of unification \cite{horava_witten1,horava_witten2} have boosted interest in higher dimensional cosmologies.  
An exciting new area of this field involves the realization, in braneworld approaches, that extra coordinates need not be compact \cite{arkani-hamed,antoniadas}. In the Randall-Sundrum model \cite{randall_sundrum1,randall_sundrum2}, for example, our observed universe, as seen at low-energy scales, is localized to four dimensions by a 3-brane embedded within a five dimensional bulk.  An interesting question concerns the range of geometries that would support such localization.  These are characterized by a warp factor, dependent in general on the fifth coordinate, that determines the geometry of the 4d part of the metric \cite{karch,felder}.  The 5D Einstein equations, along with boundary conditions, set the form of this warp factor.

Induced matter theory, proposed by Wesson\cite{wesson1}, provides another promising non-compactified higher dimensional approach, albeit in the classical realm. It postulates that the additional geometrical terms arising from the fifth coordinate in the five-dimensional vacuum Einstein equations can be associated with the matter-energy components of four-dimensional theory.\cite{wesson2} In other words, the dynamics of the observed 4D matter- and energy-filled universe are equivalent to that of a 5D empty Kaluza-Klein theory--rendering material from pure geometry. Researchers have found, for instance, that the known equations of state for radiation-dominated and matter-dominated stages of the universe arise as consequences of this model, as applied to the Robertson-Walker metric extended by an extra dimension.\cite{poncedeleon1,overduin,wesson3} A number of exact 5D solutions have been found \cite{fukui,liu1,wesson4,wesson5,wesson6,wesson7,coley,abolghasem,abolghasem2}, with cosmological implications including a variable cosmological constant and a non-singular big bounce\cite{liu2}.  A close connection has been found between this space-time-matter (STM) approach and braneworld theories\cite{poncedeleon2}.
 
If large higher dimensions exist, why don't we perceive them directly? In braneworld approaches, only gravitational fields (being closed strings) are free to move through extra dimensions, the other particles (open strings) are confined to our three dimensional brane. Therefore standard means of observation, based on the detection of electromagnetic radiation, do not detect the extra dimensions.

In a universe with non-compact extra dimensions, one can examine the dynamics of all the cosmological scale factors within the context of the higher dimensional Einstein equations.  There is no a priori reason to assume that each of these factors behaves in the same manner.  In fact, while some of these (representing conventional space) expand, others might contract.  Thus to understand these dynamics, it seems appropriate to consider higher dimensional anisotropic cosmologies.

A number of papers have examined the behavior of higher dimensional anisotropic models, considering whether or not their extra scale factors shrink down over time.\cite{chodos,freund,barrow,furusawa,demianski1,demianski2,biesada,demaret,halpern1}  We found such dimensional reduction in the case of a five dimensional extension of the Kasner solution within the context of induced matter theory \cite{halpern2}.  The requirement of positive density led naturally to the contraction over time of the extra scale factor.  Recently, Frolov has obtained a class of exact anisotropic solutions within the context of the brane world approach, in which the behavior of the bulk resembles the Kasner solution \cite{frolov}.  

The Kasner solution represents the simplest type of anisotropic behavior.  As Belinskii, Khalatnikov and Lifshitz have shown, the very early dynamics of the universe may have been far more complex \cite{bkl}.  The range of possible anisotropic geometries can be characterized by Bianchi's classification of the isometry groups of three dimensional Riemannian manifolds into nine distinct types (I-IX) \cite{bianchi}.  This work was first applied to cosmology by Taub in 1951 \cite{taub}.  Calculating the field equations for each Bianchi type, Taub examined spatially homogeneous, vacuum-filled space times, classifying the Kasner solution as Bianchi type-I and finding a new exact solution in the case of type-II.

To extend these results to higher dimensions and understand the range of possible behaviors, we have examined several different anisotropic five dimensional cosmologies, including generalized versions of type-II and type-V, comparing these solutions to our previous results for type-I.  Ultimately, our aim is to help answer the question: of the range of solutions to the five dimensional Einstein equations (including the full variety of anisotropic geometries), which of these exhibit long term behavior consistent with observed present-day conditions? 

\section{5D Extensions of Taub's Bianchi type-II Solution}
\label{sec:bianchi-II}

We consider the five dimensional metric:

\begin{equation}
\label{eqn:metric}
ds^2 = e^{\nu} dt^2 - g_{ij} {\omega^i}{\omega^j} - e^{\mu} d{\l}^2 \nonumber
\end{equation}

where the 3D spatial part of the metric can be expressed in diagonal form as:
\begin{equation}
g_{ij} = diag(e^{\alpha},e^{\beta},e^{\gamma})
\end{equation}

The time coordinate $t$ and the three spatial coordinates $x$, $y$ and $z$ have been supplemented with a fifth coordinate $\l$.  We assume that the metric coefficients $\mu$, $\nu$, $\alpha$, $\beta$ and $\gamma$ each depend, in general, on both $t$ and $\l$. 

The one-forms ${\omega^i}$ have the relationship:
\begin{equation}
d{\omega^i} = \frac{1}{2} {C^i}_{jk} {\omega^j}{\omega^k}
\end{equation}
where the ${C^i}_{jk}$ are the structure constants corresponding to the particular Bianchi type.  

In the case of type-II, the non-zero structure constants are: 
\begin{equation}
{C^1}_{23}= -{C^1}_{32}= 1
\end{equation}

The 5D Einstein tensor for the type-II case can be written in the form:
\begin{eqnarray}
\label{eqn:ein1}
G^0_0= e^{-\nu} ( & & -\frac{1}{4}{{\mu}^.}{{\alpha}^.}-\frac{1}{4}{{\mu}^.}{{\beta}^.}-\frac{1}{4}{{\mu}^.}{{\gamma}^.}-\frac{1}{4}{{\alpha}^.}{{\beta}^.}-\frac{1}{4}{{\alpha}^.}{{\gamma}^.}-\frac{1}{4}{{\beta}^.}{{\gamma}^.})+ \nonumber \\ e^{-\mu} ( & &\frac{1}{2} {\alpha}^{**} +\frac{1}{2} {\beta}^{**}+ \frac{1}{2} {\gamma}^{**} + \frac{1}{4} {{\alpha}^*}^2 + \frac{1}{4} {{\beta}^*}^2 + \frac{1}{4} {{\gamma}^*}^2- \nonumber \\& & \frac{1}{4}{{\mu}^*}{{\alpha}^*} -\frac{1}{4}{{\mu}^*}{{\beta}^*}-\frac{1}{4}{{\mu}^*}{{\gamma}^*}+\frac{1}{4}{{\alpha}^*}{{\beta}^*}+\frac{1}{4}{{\alpha}^*}{{\gamma}^*}+\frac{1}{4}{{\beta}^*}{{\gamma}^*}) + \frac{e^{\alpha}}{4 e^{\beta} e^{\gamma}} \\ \nonumber \\ 
\label{eqn:ein2}
G^0_4= e^{-\nu} ( & & 2{\alpha^{.*}} + 2{\beta^{.*}} + 2{\gamma^{.*}} +{{\alpha}^.}{{\alpha}^*} + {{\beta}^.}{{\beta}^*} + {{\gamma}^.}{{\gamma}^*} - \nonumber \\ & & {{\alpha}^.}{{\nu}^*}- {{\beta}^.}{{\nu}^*}- {{\gamma}^.}{{\nu}^*} - {{\alpha}^*}{{\mu}^.}- {{\beta}^*}{{\mu}^.}- {{\gamma}^*}{{\mu}^.})\\ \nonumber \\ 
\label{eqn:ein3}
G^1_1=   e^{-\nu} ( & & -\frac{1}{2}{{\beta}^{..}}-\frac{1}{2}{{\gamma}^{..}}-\frac{1}{2}{{\mu}^{..}}-\frac{1}{4}{{\beta}^.}^2-\frac{1}{4}{{\gamma}^.}^2-\frac{1}{4}{{\mu}^.}^2- \nonumber \\ & & \frac{1}{4}{{\beta}^.}{{\gamma}^.} -\frac{1}{4}{{\beta}^.}{{\mu}^.} - \frac{1}{4}{{\gamma}^.}{{\mu}^.} + \frac{1}{4}{{\beta}^.}{{\nu}^.} +\frac{1}{4}{{\gamma}^.}{{\nu}^.} + \frac{1}{4}{{\mu}^.}{{\nu}^.}) + \nonumber \\ e^{-\mu} ( & & \frac{1}{2}{{\beta}^{**}} + \frac{1}{2}{{\gamma}^{**}} + \frac{1}{2}{{\nu}^{**}} + \frac{1}{4}{{\beta}^*}^2 +\frac{1}{4}{{\gamma}^*}^2 + \frac{1}{4}{{\nu}^*}^2+ \nonumber \\ & & \frac{1}{4} {{\beta}^*}{{\gamma}^*} + \frac{1}{4} {{\beta}^*}{{\nu}^*} +\frac{1}{4} {{\gamma}^*}{{\nu}^*} - \frac{1}{4}{{\beta}^*}{{\mu}^*} - \frac{1}{4}{{\gamma}^*}{{\mu}^*} - \frac{1}{4}{{\mu}^*}{{\nu}^*})+ \frac{3 e^{\alpha}}{4 e^{\beta} e^{\gamma}} \\ \nonumber \\
\label{eqn:ein4}
G^2_2=   e^{-\nu} ( & & -\frac{1}{2}{{\alpha}^{..}}-\frac{1}{2}{{\gamma}^{..}}-\frac{1}{2}{{\mu}^{..}}-\frac{1}{4}{{\alpha}^.}^2-\frac{1}{4}{{\gamma}^.}^2-\frac{1}{4}{{\mu}^.}^2- \nonumber \\ & & \frac{1}{4}{{\alpha}^.}{{\gamma}^.} -\frac{1}{4}{{\alpha}^.}{{\mu}^.} - \frac{1}{4}{{\gamma}^.}{{\mu}^.} + \frac{1}{4}{{\alpha}^.}{{\nu}^.} +\frac{1}{4}{{\gamma}^.}{{\nu}^.} + \frac{1}{4}{{\mu}^.}{{\nu}^.}) + \nonumber \\ e^{-\mu} ( & & \frac{1}{2}{{\alpha}^{**}} + \frac{1}{2}{{\gamma}^{**}} + \frac{1}{2}{{\nu}^{**}} + \frac{1}{4}{{\alpha}^*}^2 +\frac{1}{4}{{\gamma}^*}^2 + \frac{1}{4}{{\nu}^*}^2+ \nonumber \\ & & \frac{1}{4} {{\alpha}^*}{{\gamma}^*} + \frac{1}{4} {{\alpha}^*}{{\nu}^*} +\frac{1}{4} {{\gamma}^*}{{\nu}^*} - \frac{1}{4}{{\alpha}^*}{{\mu}^*} - \frac{1}{4}{{\gamma}^*}{{\mu}^*} - \frac{1}{4}{{\mu}^*}{{\nu}^*}) - \frac{e^{\alpha}}{4 e^{\beta} e^{\gamma}}\\ \nonumber \\
\label{eqn:ein5}
G^3_3=   e^{-\nu} ( & & -\frac{1}{2}{{\alpha}^{..}}-\frac{1}{2}{{\beta}^{..}}-\frac{1}{2}{{\mu}^{..}}-\frac{1}{4}{{\alpha}^.}^2-\frac{1}{4}{{\beta}^.}^2-\frac{1}{4}{{\mu}^.}^2- \nonumber \\ & & \frac{1}{4}{{\alpha}^.}{{\beta}^.} -\frac{1}{4}{{\alpha}^.}{{\mu}^.} - \frac{1}{4}{{\beta}^.}{{\mu}^.} + \frac{1}{4}{{\alpha}^.}{{\nu}^.} +\frac{1}{4}{{\beta}^.}{{\nu}^.} + \frac{1}{4}{{\mu}^.}{{\nu}^.}) + \nonumber \\ e^{-\mu} ( & & \frac{1}{2}{{\alpha}^{**}} + \frac{1}{2}{{\beta}^{**}} + \frac{1}{2}{{\nu}^{**}} + \frac{1}{4}{{\alpha}^*}^2 +\frac{1}{4}{{\beta}^*}^2 + \frac{1}{4}{{\nu}^*}^2+ \nonumber \\ & & \frac{1}{4} {{\alpha}^*}{{\beta}^*} + \frac{1}{4} {{\alpha}^*}{{\nu}^*} +\frac{1}{4} {{\beta}^*}{{\nu}^*} - \frac{1}{4}{{\alpha}^*}{{\mu}^*} - \frac{1}{4}{{\beta}^*}{{\mu}^*} - \frac{1}{4}{{\mu}^*}{{\nu}^*}) - \frac{e^{\alpha}}{4 e^{\beta} e^{\gamma}}\\ \nonumber \\
\label{eqn:ein6}
G^4_4=   e^{-\nu} ( & & -\frac{1}{2}{{\alpha}^{..}} -\frac{1}{2}{{\beta}^{..}}-\frac{1}{2}{{\gamma}^{..}} -\frac{1}{4}{{\alpha}^.}^2-\frac{1}{4}{{\beta}^.}^2-\frac{1}{4}{{\gamma}^.}^2- \nonumber \\ & & \frac{1}{4}{{\alpha}^.}{{\beta}^.} -\frac{1}{4}{{\alpha}^.}{{\gamma}^.} - \frac{1}{4}{{\beta}^.}{{\gamma}^.} + \frac{1}{4}{{\alpha}^.}{{\nu}^.} +\frac{1}{4}{{\beta}^.}{{\nu}^.} + \frac{1}{4}{{\gamma}^.}{{\nu}^.}) + \nonumber \\ e^{-\mu} ( & & \frac{1}{4} {{\alpha}^*}{{\beta}^*} + \frac{1}{4} {{\alpha}^*}{{\gamma}^*} +\frac{1}{4} {{\beta}^*}{{\gamma}^*} -\frac{1}{4} {{\alpha}^*}{{\nu}^*} + \frac{1}{4} {{\beta}^*}{{\nu}^*} + \frac{1}{4}{{\gamma}^*}{{\nu}^*})+ \frac{e^{\alpha}}{4 e^{\beta} e^{\gamma}}
\end{eqnarray}

where we use overdots to represent partial derivatives with respect to $t$, and asterisks to represent partial derivatives with respect to $\l$.

Following Wesson's procedure we can rewrite the 5D vacuum Einstein equations, ${G^{\mu}}_{\nu} = 0$ as 4D equations with induced matter by collecting each of the terms in $G^0_0$, $G^1_1$, $G^2_2$ and $G^3_3$  dependent on either $\mu$ or on derivatives with respect to $l$, placing these quantities on the right-hand side of the equations and identifying them respectively as the induced matter density and pressure components:

\begin{eqnarray}
\label{eqn:density}
8 \pi \rho= e^{-\nu} ( & & -\frac{1}{4} {{\alpha}^.}{{\mu}^.}- \frac{1}{4} {{\beta}^.}{{\mu}^.}- \frac{1}{4} {{\gamma}^.}{{\mu}^.}) + \nonumber \\ e^{-\mu} ( & & \frac{1}{2} {\alpha}^{**} +\frac{1}{2} {\beta}^{**}+ \frac{1}{2} {\gamma}^{**} + \frac{1}{4} {{\alpha}^*}^2 + \frac{1}{4} {{\beta}^*}^2 + \frac{1}{4} {{\gamma}^*}^2- \nonumber \\ & & \frac{1}{4}{{\mu}^*}{{\alpha}^*}-\frac{1}{4}{{\mu}^*}{{\beta}^*}-\frac{1}{4}{{\mu}^*}{{\gamma}^*}+\frac{1}{4}{{\alpha}^*}{{\beta}^*}+\frac{1}{4}{{\alpha}^*}{{\gamma}^*}+\frac{1}{4}{{\beta}^*}{{\gamma}^*})\\ \nonumber \\
\label{eqn:press1}
8 \pi p_1 = e^{-\nu} ( & & \frac{1}{2}{{\mu}^{..}}+ \frac{1}{4}{{\mu}^.}^2+ \frac{1}{4} {{\beta}^.}{{\mu}^.}+ \frac{1}{4} {{\gamma}^.}{{\mu}^.}- \frac{1}{4} {{\mu}^.}{{\nu}^.}) + \nonumber \\ e^{-\mu} ( & & -\frac{1}{2}{{\beta}^{**}} - \frac{1}{2}{{\gamma}^{**}} -\frac{1}{2}{{\nu}^{**}} - \frac{1}{4}{{\beta}^*}^2 -\frac{1}{4}{{\gamma}^*}^2 - \frac{1}{4}{{\nu}^*}^2- \nonumber \\ & & \frac{1}{4} {{\beta}^*}{{\gamma}^*} - \frac{1}{4} {{\beta}^*}{{\nu}^*} -\frac{1}{4} {{\gamma}^*}{{\nu}^*} + \frac{1}{4}{{\beta}^*}{{\mu}^*} + \frac{1}{4}{{\gamma}^*}{{\mu}^*} + \frac{1}{4}{{\mu}^*}{{\nu}^*})\\ \nonumber \\
\label{eqn:press2}
8 \pi p_2 = e^{-\nu} ( & & \frac{1}{2}{{\mu}^{..}}+ \frac{1}{4}{{\mu}^.}^2+ \frac{1}{4} {{\alpha}^.}{{\mu}^.}+ \frac{1}{4} {{\gamma}^.}{{\mu}^.}- \frac{1}{4} {{\mu}^.}{{\nu}^.}) +\nonumber \\ e^{-\mu} ( & & -\frac{1}{2}{{\alpha}^{**}} - \frac{1}{2}{{\gamma}^{**}} - \frac{1}{2}{{\nu}^{**}} - \frac{1}{4}{{\alpha}^*}^2 -\frac{1}{4}{{\gamma}^*}^2 - \frac{1}{4}{{\nu}^*}^2- \nonumber \\ & & \frac{1}{4} {{\alpha}^*}{{\gamma}^*} - \frac{1}{4} {{\alpha}^*}{{\nu}^*} -\frac{1}{4} {{\gamma}^*}{{\nu}^*} + \frac{1}{4}{{\alpha}^*}{{\mu}^*} + \frac{1}{4}{{\gamma}^*}{{\mu}^*} + \frac{1}{4}{{\mu}^*}{{\nu}^*})\\ \nonumber \\
\label{eqn:press3}
8 \pi p_3 = e^{-\nu} ( & & \frac{1}{2}{{\mu}^{..}}+ \frac{1}{4}{{\mu}^.}^2+ \frac{1}{4} {{\alpha}^.}{{\mu}^.}+ \frac{1}{4} {{\beta}^.}{{\mu}^.}- \frac{1}{4} {{\mu}^.}{{\nu}^.}) + \nonumber \\ e^{-\mu} ( & & -\frac{1}{2}{{\alpha}^{**}} - \frac{1}{2}{{\beta}^{**}} - \frac{1}{2}{{\nu}^{**}} - \frac{1}{4}{{\alpha}^*}^2 -\frac{1}{4}{{\beta}^*}^2 - \frac{1}{4}{{\nu}^*}^2- \nonumber \\ & & \frac{1}{4} {{\alpha}^*}{{\beta}^*} - \frac{1}{4} {{\alpha}^*}{{\nu}^*} -\frac{1}{4} {{\beta}^*}{{\nu}^*} + \frac{1}{4}{{\alpha}^*}{{\mu}^*} + \frac{1}{4}{{\beta}^*}{{\mu}^*} + \frac{1}{4}{{\mu}^*}{{\nu}^*})
\end{eqnarray}

Note that definitions (\ref{eqn:density} - \ref{eqn:press3}) contain no references to terms dependent on particular structure constants and are therefore generally true for all anisotropic models with metrics of the form (\ref{eqn:metric}).

In the case of type-II the Einstein equations now read:

\begin{eqnarray}
\label{eqn:einstein0}
e^{-\nu} ( & & -\frac{1}{4}{{\alpha}^.}{{\beta}^.}-\frac{1}{4}{{\alpha}^.}{{\gamma}^.}-\frac{1}{4}{{\beta}^.}{{\gamma}^.})+ \frac{e^{\alpha}}{4 e^{\beta} e^{\gamma}}+ 8 \pi \rho = 0\\ \nonumber \\ 
\label{eqn:einstein1}
e^{-\nu} ( & & -\frac{1}{2}{{\beta}^{..}}-\frac{1}{2}{{\gamma}^{..}}-\frac{1}{4}{{\beta}^.}^2-\frac{1}{4}{{\gamma}^.}^2- \frac{1}{4}{{\beta}^.}{{\gamma}^.} + \frac{1}{4}{{\beta}^.}{{\nu}^.} +\frac{1}{4}{{\gamma}^.}{{\nu}^.})+ \frac{3 e^{\alpha}}{4 e^{\beta} e^{\gamma}} - 8 \pi p_1 = 0 \\ \nonumber \\
\label{eqn:einstein2}
e^{-\nu} ( & & -\frac{1}{2}{{\alpha}^{..}}-\frac{1}{2}{{\gamma}^{..}}- \frac{1}{4}{{\alpha}^.}^2-\frac{1}{4}{{\gamma}^.}^2- \frac{1}{4}{{\alpha}^.}{{\gamma}^.}  + \frac{1}{4}{{\alpha}^.}{{\nu}^.} +\frac{1}{4}{{\gamma}^.}{{\nu}^.}) - \frac{e^{\alpha}}{4 e^{\beta} e^{\gamma}} - 8 \pi p_2 = 0 \\ \nonumber \\ 
\label{eqn:einstein3}
e^{-\nu} ( & & -\frac{1}{2}{{\alpha}^{..}}-\frac{1}{2}{{\beta}^{..}}-\frac{1}{4}{{\alpha}^.}^2-\frac{1}{4}{{\beta}^.}^2- \frac{1}{4}{{\alpha}^.}{{\beta}^.} + \frac{1}{4}{{\alpha}^.}{{\nu}^.} +\frac{1}{4}{{\beta}^.}{{\nu}^.}) - \frac{e^{\alpha}}{4 e^{\beta} e^{\gamma}} - 8 \pi p_3 = 0 \\ \nonumber \\
\label{eqn: einstein4}
e^{-\nu} ( & & 2{\alpha^{.*}} + 2{\beta^{.*}} + 2{\gamma^{.*}} +{{\alpha}^.}{{\alpha}^*} + {{\beta}^.}{{\beta}^*} + {{\gamma}^.}{{\gamma}^*} - \nonumber \\ & & {{\alpha}^.}{{\nu}^*}- {{\beta}^.}{{\nu}^*}- {{\gamma}^.}{{\nu}^*} - {{\alpha}^*}{{\mu}^.}- {{\beta}^*}{{\mu}^.}- {{\gamma}^*}{{\mu}^.}) = 0\\ \nonumber \\ 
\label{eqn:einstein5}
e^{-\nu} ( & & -\frac{1}{2}{{\alpha}^{..}} -\frac{1}{2}{{\beta}^{..}}-\frac{1}{2}{{\gamma}^{..}} -\frac{1}{4}{{\alpha}^.}^2-\frac{1}{4}{{\beta}^.}^2-\frac{1}{4}{{\gamma}^.}^2- \nonumber \\ & & \frac{1}{4}{{\alpha}^.}{{\beta}^.} -\frac{1}{4}{{\alpha}^.}{{\gamma}^.} - \frac{1}{4}{{\beta}^.}{{\gamma}^.} + \frac{1}{4}{{\alpha}^.}{{\nu}^.} +\frac{1}{4}{{\beta}^.}{{\nu}^.} + \frac{1}{4}{{\gamma}^.}{{\nu}^.}) + \nonumber \\ e^{-\mu} ( & & \frac{1}{4} {{\alpha}^*}{{\beta}^*} + \frac{1}{4} {{\alpha}^*}{{\gamma}^*} +\frac{1}{4} {{\beta}^*}{{\gamma}^*} +\frac{1}{4} {{\alpha}^*}{{\nu}^*} + \frac{1}{4} {{\beta}^*}{{\nu}^*} + \frac{1}{4}{{\gamma}^*}{{\nu}^*}) + \frac{e^{\alpha}}{4 e^{\beta} e^{\gamma}}
 = 0
\end{eqnarray} 

Solving the 5D Einstein equations for the vacuum case, we find two distinct sets of solutions.
The first class of solutions is independent of $l$ and represents an extension of Taub's 4D solution.  It can be stated as:

\begin{eqnarray}
\alpha &=& 2{a_1}t - \ln{[\cosh{(t+t_0)}]}\\
\beta &=& 2{b_1}t + \ln{[\cosh{(t+t_0)}]}\\
\gamma &=& 2{c_1}t + \ln{[\cosh{(t+t_0)}]}\\
\mu &=& -4{a_1}t\\ 
\nu &=& -4{a_1}t + 4{b_1}t + 2\ln{[\cosh{(t+t_0)}]}
\end{eqnarray}

where:
\begin{equation}
c_1= \frac{8 {a_1}^2 + 4 a_1 b_1 +1}{4 b_1 - 4 a_1}
\end{equation}

The density and pressure of the associated induced matter can be calculated by substituting this solution into eqns. (\ref{eqn:density}-\ref{eqn:press3}).  One obtains:
\begin{eqnarray}
8 \pi \rho &=& a_1 [\tanh(t+t_0) + \frac{(4{a_1}^2 + 4a_1 b_1 + 4{b_1}^2 + 1)}{2(b_1 - a_1)}] e^{f(t)}\\
8 \pi p_1 &=& a_1 [\frac{(8{a_1}^2 + 4{b_1}^2 + 1)}{2(b_1 - a_1)}]e^{f(t)}\\
8 \pi p_2 &=& a_1 [\tanh(t+t_0) + \frac{(12{a_1}^2 - 8 a_1 b_1 + 8 {b_1}^2 + 1)}{2(b_1 - a_1)}]e^{f(t)}\\
8 \pi p_3 &=& a_1 [2 \tanh(t+t_0) + \frac{(10{a_1}^2 + 2{b_1}^2 + 1)}{(b_1 - a_1)}]e^{f(t)}\\
\nonumber
\end{eqnarray}

where:
\begin{equation}
f(t)=\frac{(12{a_1}^2 - 4a_1 b_1 + 4{b_1}^2 + 1) t}{(a_1-b_1)} - 2 \ln{[\cosh{(t+t_0)}]}
\end{equation}

To ensure that the density of the induced matter is strictly positive, a physically realistic condition, we find that $a_1 > 0$ and $b_1 > a_1$.  This first condition mandates that $\mu < 0$ for all $t$, forcing the scale factor associated with the fifth coordinate to be a monotonically decreasing function.  Thus the positive density requirement naturally leads to dimensional reduction.  This range of values for $a_1$ and $b_1$ also ensures that $c_1$ is positive as well, guaranteeing that the three spatial scale factors expand over time.  Hence one ends up in the long term with a situation in which three of the dimensions are large and one is small, a possible model of present-day conditions.

\section{A Class of 5D Bianchi Type-II Solutions with Exponential Behavior}

Further investigating eqns (\ref{eqn:einstein0} - \ref{eqn:einstein5}) we find a second set of five dimensional solutions with type-II geometry.  In contrast to the first set, this
class of solutions is dependent on both $t$ and $l$.  The metric coefficients can be expressed in the following manner:
\begin{eqnarray}
\alpha &=& 2{a_1}t + 2{a_2}l\\
\beta &=& 2{b_1}t + 2{b_2}l\\
\gamma &=& 2{c_1}t + 2{c_2}l\\
\mu &=& \nu = 2{d_1}t + 2{d_2}l
\end{eqnarray}

where:

\begin{eqnarray}
a_1 &=& {\frac {-2\,{a_2}^{2}+2\,a_2\,c_2+1+a_2\,
\sqrt {6}\sqrt {2\,{a_2}^{2}+2\,{c_2}^{2}-1}}{2\sqrt {4\,{{
a_2}}^{2}-1-2\,a_2\,c_2+4\,{c_2}^{2}+\sqrt {6}\sqrt 
{\left (2\,{a_2}^{2}+2\,{c_2}^{2}-1\right )\left (a_2-{
c_2}\right )^{2}}}}}
\\
b_1 &=& {\frac {-6\,{a_2}^{2}+4\,a_2\,c_2-10\,{c_2}^{
2}+2+{(a_2-3c_2)}\,\sqrt {6}\sqrt {2\,{a_2}^{2}+2\,{c_2}^{2}-1}}{2
\sqrt {4\,{a_2}^{2}-1-2\,a_2\,c_2+4\,{c_2}^{2}+
\sqrt {6}\sqrt {\left (2\,{a_2}^{2}+2\,{c_2}^{2}-1\right )
\left (a_2-c_2\right )^{2}}}}}
\\
c_1 &=& {\frac {-2\,a_2\,c_2-1+2\,{c_2}^{2}+c_2\sqrt {6}
\sqrt {2\,{a_2}^{2}+2\,{c_2}^{2}-1}}{2\sqrt {4\,{{
a_2}}^{2}-1-2\,a_2\,c_2+4\,{c_2}^{2}+\sqrt {6}\sqrt 
{\left (2\,{a_2}^{2}+2\,{c_2}^{2}-1\right )\left (a_2-{
c_2}\right )^{2}}}}}
\\
d_1 &=& -{\frac {2\,{a_2}^{2}+4\,{c_2}^{2}+c_2\sqrt {6}\sqrt {2\,{{\it 
a2}}^{2}+2\,{c_2}^{2}-1}}{\sqrt {4\,{a_2}^{2}-1-2\,{
a_2}\,c_2+4\,{c_2}^{2}+\sqrt {6}\sqrt {\left (2\,{{\it a2
}}^{2}+2\,{c_2}^{2}-1\right )\left (a_2-c_2\right )^{2}
}}}}
\\
b_2 &=& -2c_2 - \frac{\sqrt{6}}{2} \sqrt {2 {a_2}^2 + 2 {c_2}^2 - 1}
\\
d_2 &=& -a_2- c_2 - \frac{\sqrt{6}}{2} \sqrt {2 {a_2}^2 + 2 {c_2}^2 - 1}
\end{eqnarray}

and $a_2$ and $c_2$ are independent parameters with $c_2 > a_2$.

Interestingly, this set of solutions is purely exponential in character, with monotonic behavior similar to the Kasner (type-I) solution.  Note, however, that the relationship amongst these exponents is more complex than in the Kasner case.  Also, unlike the 5D Kasner solution, in this case is it impossible for three of the scale factors to behave identically.  Therefore this type-II model cannot expand isotropically in three directions while contracting in the fourth. 

For this class of solutions the density and pressure of the induced matter can be calculated to be:

\begin{eqnarray}
8 \pi \rho &=& [-2\,{{c_2}}^{2}-{a_2}\,{c_2}-1/2\,\left ({a_2}+{c_2}
\right )\sqrt {6}\sqrt {2\,{{a_2}}^{2}+2\,{{c_2}}^{2}-1}] e^{{\kappa_1}t+{\kappa_2}l}\\ \nonumber \\
8 \pi p_1 &=& [-14\,{{c_2}}^{4}+20\,{a_2}\,{{c_2}}^{3}+13\,{{c_2}}^{2}-22
\,{{c_2}}^{2}{{a_2}}^{2} \nonumber \\
&-&9\,{c_2}\,{a_2}+16\,{c_2}\,{{
a_2}}^{3}-3/2-6\,{{a_2}}^{4}+8\,{{a_2}}^{2} \nonumber \\
&+&(5/2\,{c_2}-3/2\,{a_2}-4\,{{c_2}}^{3}-4\,{{a_2}}^{2}{c_2
}+6\,{{c_2}}^{2}{a_2} \nonumber \\
&+&2\,{{a_2}}^{3})\sqrt {6}\sqrt {2\,{a_2}^{2}+2\,{c_2}^{2}-1}] e^{{\kappa_1}t+ {\kappa_2}l} \\ \nonumber \\
8 \pi p_2 &=& [-14\,{{c_2}}^{4}-8\,{a_2}\,{{c_2}}^{3}-10\,{{c_2}}^{2}{{
a_2}}^{2}+4\,{{c_2}}^{2} \nonumber \\&-&8\,{c_2}\,{{a_2}}^{3}+4\,{c_2}
\,{a_2}-2\,{{a_2}}^{4}+{{a_2}}^{2} \nonumber \\
&-&(4{c_2}^3+2{a_2}{c_2}^2)\sqrt {6}\sqrt {2\,{a_2}^{2}+2\,{c_2}^{2}-1}]e^{{\kappa_1}t+ {\kappa_2}l}\\ \nonumber \\
8 \pi p_3 &=& [-14\,{{\it a2}}^{4}+28\,{\it c2}\,{{\it a2}}^{3}-34\,{{\it c2}}^{2}{{
\it a2}}^{2}+10\,{{\it a2}}^{2} \nonumber \\&+&28\,{\it a2}\,{{\it c2}}^{3}-14\,{\it 
c2}\,{\it a2}-14\,{{\it c2}}^{4}+10\,{{\it c2}}^{2}-3/2 \nonumber \\
&-&(4{a_2}^2-2{a_2}{c_2}+2{c_2}^2-1)\sqrt{6}\sqrt{(2{a_2}^2+2{c_2}^2-1)({c_2}-{a_2})^2}]e^{{\kappa_1}t+ {\kappa_2}l}/
 \nonumber \\
&[& 4{a_2}^2-2{a_2}{c_2}+4*{c_2}^2-1+ \sqrt{6} \sqrt{2 {a_2}^2 + 2 {c_2}^2 - 1}]  \\ \nonumber
\end{eqnarray}

where:

\begin{eqnarray} 
\kappa_1 &=& {\frac {8\,{{c_2}}^{2}+4\,{{a_2}}^{2}+2\,{c_2}\,\sqrt {6}
\sqrt {2\,{{a_2}}^{2}+2\,{{c_2}}^{2}-1}}{\sqrt {4\,{{c_2}}^{2
}-2\,{a_2}\,{c_2}+4\,{{a_2}}^{2}-1+\sqrt {6}\sqrt {\left (2\,
{{a_2}}^{2}+2\,{{c_2}}^{2}-1\right )\left ({c_2}-{a_2}
\right )^{2}}}}}\\
\kappa_2 &=& 2(a_2+c_2) +\sqrt{6} \sqrt {2\,{a_2}^{2}+2\,{c_2}^{2}-1}
\end{eqnarray}

To guarantee the physically realistic requirement of positive density, one can chose $c_2 < 0$.

\section{5D BIANCHI TYPE-V SOLUTION} 

We now consider 5D models with Bianchi type-V spatial geometry. We write the metric in the same manner as (\ref{eqn:metric}) with the non-zero structure constants of the Lie algebra of the one-forms equal to:

\begin{eqnarray}
{C^1}_{13}&=& -{C^1}_{31}= 1 \\
{C^2}_{23}&=& -{C^2}_{32}= 1 \\
\nonumber
\end{eqnarray}

The Einstein equations can be written as:

\begin{eqnarray}
\label{eqn:einst0}
e^{-\nu} ( & & -\frac{1}{4}{{\alpha}^.}{{\beta}^.}-\frac{1}{4}{{\alpha}^.}{{\gamma}^.}-\frac{1}{4}{{\beta}^.}{{\gamma}^.})+ {e^{-\gamma}}+ 8 \pi \rho = 0\\ \nonumber \\ 
\label{eqn:einst1}
e^{-\nu} ( & & -\frac{1}{2}{{\beta}^{..}}-\frac{1}{2}{{\gamma}^{..}}-\frac{1}{4}{{\beta}^.}^2-\frac{1}{4}{{\gamma}^.}^2- \frac{1}{4}{{\beta}^.}{{\gamma}^.} + \frac{1}{4}{{\beta}^.}{{\nu}^.} +\frac{1}{4}{{\gamma}^.}{{\nu}^.})+ {e^{-\gamma}} + 8 \pi p_1 = 0 \\ \nonumber \\
\label{eqn:einst2}
e^{-\nu} ( & & -\frac{1}{2}{{\alpha}^{..}}-\frac{1}{2}{{\gamma}^{..}}- \frac{1}{4}{{\alpha}^.}^2-\frac{1}{4}{{\gamma}^.}^2- \frac{1}{4}{{\alpha}^.}{{\gamma}^.}  + \frac{1}{4}{{\alpha}^.}{{\nu}^.} +\frac{1}{4}{{\gamma}^.}{{\nu}^.}) + {e^{-\gamma}} + 8 \pi p_2 = 0 \\ \nonumber \\ 
\label{eqn:einst3}
e^{-\nu} ( & & -\frac{1}{2}{{\alpha}^{..}}-\frac{1}{2}{{\beta}^{..}}-\frac{1}{4}{{\alpha}^.}^2-\frac{1}{4}{{\beta}^.}^2- \frac{1}{4}{{\alpha}^.}{{\beta}^.} + \frac{1}{4}{{\alpha}^.}{{\nu}^.} +\frac{1}{4}{{\beta}^.}{{\nu}^.}) - {e^{-\gamma}} + 8 \pi p_3 = 0 \\ \nonumber \\
\label{eqn: einst4}
e^{-\nu} ( & & 2{\alpha^{.*}} + 2{\beta^{.*}} + 2{\gamma^{.*}} +{{\alpha}^.}{{\alpha}^*} + {{\beta}^.}{{\beta}^*} + {{\gamma}^.}{{\gamma}^*} - \nonumber \\ & & {{\alpha}^.}{{\nu}^*}- {{\beta}^.}{{\nu}^*}- {{\gamma}^.}{{\nu}^*} - {{\alpha}^*}{{\mu}^.}- {{\beta}^*}{{\mu}^.}- {{\gamma}^*}{{\mu}^.}) = 0\\ \nonumber \\ 
\label{eqn:einst5}
e^{-\nu} ( & & -\frac{1}{2}{{\alpha}^{..}} -\frac{1}{2}{{\beta}^{..}}-\frac{1}{2}{{\gamma}^{..}} -\frac{1}{4}{{\alpha}^.}^2-\frac{1}{4}{{\beta}^.}^2-\frac{1}{4}{{\gamma}^.}^2- \nonumber \\ & & \frac{1}{4}{{\alpha}^.}{{\beta}^.} -\frac{1}{4}{{\alpha}^.}{{\gamma}^.} - \frac{1}{4}{{\beta}^.}{{\gamma}^.} + \frac{1}{4}{{\alpha}^.}{{\nu}^.} +\frac{1}{4}{{\beta}^.}{{\nu}^.} + \frac{1}{4}{{\gamma}^.}{{\nu}^.}) + \nonumber \\ e^{-\mu} ( & & \frac{1}{4} {{\alpha}^*}{{\beta}^*} + \frac{1}{4} {{\alpha}^*}{{\gamma}^*} +\frac{1}{4} {{\beta}^*}{{\gamma}^*} +\frac{1}{4} {{\alpha}^*}{{\nu}^*} + \frac{1}{4} {{\beta}^*}{{\nu}^*} + \frac{1}{4}{{\gamma}^*}{{\nu}^*}) + {e^{-\gamma}}
 = 0
\end{eqnarray}

where expressions for the induced density and pressure are identical to eqns. (\ref{eqn:density} - \ref{eqn:press3}).

These equations yield the following solution:

\begin{eqnarray}
\alpha &=& 2\,{a_1}\,t+\sqrt {2}{a_2}\,l\\
\beta &=& 2\,\ln (1/2\,\sqrt {-4\,{{a_1}}^{2}+2\,{{a_2}}^{2}})-2\,{a_1}
\,t-\sqrt {2}{a_2}\,l\\
\gamma &=& 2\,{a_2}\,t-2\,\ln (1/2\,\sqrt {-4\,{{a_1}}^{2}+2\,{{a_2}}^{2
}})+2\,{a_1}\,\sqrt {2}l\\
\mu &=& \nu = 2\,{a_2}\,t+2\,{a_1}\,\sqrt {2}l
\end{eqnarray}

where $a_1$ and $a_2$ are independent parameters with ${a_2}^2 > 2 {a_1}^2$ to ensure that all scale factors are real.

The density and pressure of the induced matter reduce to:
\begin{eqnarray}
8 \pi \rho &=& 8 \pi P_3 = -1/2\,\left ({a_2}^{2}\right ) {e^{-2\,{a_2}
\,t-2\,{a_1}\,\sqrt {2}l}}
\\
8 \pi P_1 &=& 8 \pi P_2 = 1/2\,\left ({{a_2}}^{2}-4\,{{a_1}}^{2}\right ){e^{-2\,{a_2}
\,t-2\,{a_1}\,\sqrt {2}l}}
\end{eqnarray}

Note that within the context of induced matter theory this solution doesn't appear to be physical because $\rho \le 0$ for all values of the parameters.

\section{Conclusions}
\label{sec:conclusions}

We have found exact solutions for several types of five dimensional anisotropic cosmologies, representing generalizations of the Bianchi type-II and type-V models.  We have interpreted these solutions within the context of induced matter theory, also known as the space-time-matter (STM) approach.  

One class of generalized Bianchi type-II cosmologies resembles an higher dimensional extension of Taub's solution.  It exhibits natural cosmological dimensional reduction, generated by the assumption of positive induced matter density.  The scale factor associated with the fifth coordinate contracts monotonically over time.  This offers a possible explanation for present-day unobservability.  The three other scale factors exhibit expansion, albeit in different manners.  Unlike the type-I case, this does not permit spatial isotropy as a special solution.  If one envisions the very early universe to have such behavior, then one must posit a mechanism for spatial isotropization (an inflationary phase, for example). 

Another class of generalized type-II models more closely resembles the monotonic exponential behavior of the Kasner model, albeit with a more complex relationship among its parameters.  It can satisfy the positive density condition for a certain parameter range.  Unlike the first class of type-II solutions, its scale factors are dependent on the fifth coordinate as well as on time.

Finally we have found a class of five dimensional cosmologies with type-V spatial geometries.  Once again, this model exhibits monotonic exponential behavior, dependent on both the fifth coordinate and time.  However, unlike the generalized type-II solutions, no range of its parameters yield positive density, violating physical intuition.

Future studies will focus on additional types of higher dimensional anisotropic models, including the effects of a negative cosmological term, considering anisotropic generalizations of five dimensional anti-de Sitter geometries.
 
\begin{acknowledgments}
We are grateful for valuable discussions with Paul S. Wesson, and for the hospitality of the physics department of the University of Waterloo.
\end{acknowledgments}

\end{document}